\documentclass[english,twocolumn,prl,nofootinbib,superscriptaddress]{revtex4}
\usepackage[T1]{fontenc}
\usepackage[latin1]{inputenc}
\usepackage{amsmath}
\usepackage{graphicx}
\usepackage{amssymb}

\makeatletter
\newcommand{\urlprefix}{}
\renewcommand{\url}[1]{}

\usepackage{lmodern} 

\newcommand{\rcsRevision}{[ RCS: $ $Revision: 1.2 $ $, 
      $ $Date: 2006-12-29 15:24:14-05 $ $]}
\newcommand{\hae}{HAE }
\newcommand{\fileInfo}{{\footnotesize \hae \rcsRevision}}

\usepackage{babel}
\makeatother
\begin{document}

\newcommand{\ket}[1]{|#1\rangle}

\newcommand{\bra}[1]{\langle#1|}

\newcommand{\braopket}[3]{\left\langle #1\left|#2\right|#3\right\rangle }

\newcommand{\braket}[2]{\langle#1|#2\rangle}

\newcommand{\bsigma}{\boldsymbol{\sigma}}

\newcommand{\brho}{\boldsymbol{\rho}}

\newcommand{\bzeta}{\boldsymbol{\zeta}}

\newcommand{\bvarepsilon}{\boldsymbol{\varepsilon}}

\newcommand{\SOsplitting}{\Delta_{0}}

\newcommand{\Egap}{E_{\mathrm{0}}}

\newcommand{\spup}{{\ket{\!\uparrow}}}

\newcommand{\spdown}{{\ket{\!\downarrow}}}

\newcommand{\spupbra}{\bra{\uparrow\!}}

\newcommand{\spdownbra}{\bra{\downarrow\!}}

\newcommand{\spupup}{\ket{\!\uparrow\uparrow}}

\newcommand{\spupdown}{\ket{\!\uparrow\downarrow}}

\newcommand{\spdownup}{\ket{\!\downarrow\uparrow}}

\newcommand{\spdowndown}{\ket{\!\downarrow\downarrow}}

\newcommand{\du}{\downarrow\uparrow}

\newcommand{\ud}{\uparrow\downarrow}

\newcommand{\upup}{\uparrow\uparrow}

\newcommand{\downdown}{\downarrow\downarrow}

\newcommand{\tRecomb}{\tau_{\mathrm{em}}}

\newcommand{\EPRdotPhoton}{\psi_{\mathrm{e}}}

\newcommand{\tDD}{t_{\mathrm{d}}}

\newcommand{\DE}{\varepsilon}

\newcommand{\DeltaEPlusMinus}{E}

\newcommand{\ZeemanSplitting}{\Delta_{\mathrm{Z}}}

\newcommand{\ZeemanDiff}{\delta_{\mathrm{Z}}}

\newcommand{\dephasingRate}{\Gamma_{\varphi}}

\newcommand{\spinDecohRate}{\Gamma_{\mathrm{s}}}

\newcommand{\noiseF}{F}

\newcommand{\errorSpinDephasing}{\beta}

\title{Quantum optical interface for gate-controlled spintronic devices }

\author{Hans-Andreas Engel}

\affiliation{Department of Physics, Harvard University, Cambridge, Massachusetts
02138}

\author{Jacob M. Taylor}

\affiliation{Department of Physics, Harvard University, Cambridge, Massachusetts
02138}

\affiliation{Department of Physics, Massachusetts Institute of Technology, Massachusetts
02139}

\author{Mikhail D. Lukin}

\affiliation{Department of Physics, Harvard University, Cambridge, Massachusetts
02138}

\author{Atac Imamo\u{g}lu}

\affiliation{Institute of Quantum Electronics, ETH-H\"onggerberg, CH-8093 Z\"urich,
Switzerland. }

\begin{abstract}
We describe an opto-electronic structure in which charge and spin
degrees of freedom in electrical gate-defined quantum dots can be
coherently coupled to light. This is achieved via electron-electron
interaction or via electron tunneling into a proximal self-assembled
quantum dot. We illustrate potential applications of this approach
by considering several quantum control techniques, including optical
read-out of gate-controlled semiconductor quantum bits and controlled
generation of entangled photon-spin pairs. 
\end{abstract}
\maketitle

The spin degree of freedom of electrons in quantum dots provides a promising
system for applications in spintronics and quantum information science~\cite{Loss97}.
Quantum dots can be defined in a two-dimensional electron gas (2DEG)
using electrical gates. Such gate-defined structures allow for a fast
electrical control, which can be effective for manipulating spin quantum
bits~\cite{PettaDD,DelftESR,EKLM,HansonRMP07}. Most importantly,
gate-defined structures can be scaled up to many-qubit systems by
using state-of-the-art nanofabrication techniques to define complex
gate geometries. However, unlike their self-assembled counterparts
\cite{T1Abstreiter,Ota2SAQDvertCoh,GammonDD,AtatureSingleSpin,AbstreiterQDMolecule06,AwschSSMeas06,AtatureSingleSpinRot},
such structures is can not be coupled to the radiation field. This
limitation results from the lack of optical transitions: while the
electron states are confined, the hole states are not. In turn, this
considerably limits the potential applicability of a diverse set of
quantum control techniques developed in AMO physics.  It is an important
consideration for scalability \cite{steane00,DuanScale04} and precludes
the use of such systems in quantum communication~\cite{briegel98,dur99,ZollerRepeater,LiliRepeater}.

This Letter describes a technique for coherently coupling spin degrees
of freedom in gate-controlled quantum dots to light. This is achieved
via either electron-electron interaction or electron tunneling into
a proximal self-assembled quantum dot (SAQD).The SAQDs can be formed
from a material with a smaller band gap than its environment and can
be addressed optically. Optical coupling to selected single spin states~\cite{T1Abstreiter,AtatureSingleSpin}
and important elements of all-optical read-out~\cite{AwschSSMeas06,AtatureSingleSpinRot}
have already been demonstrated in SAQDs. Pairs of coupled SAQDs have
been successfully fabricated and studied~\cite{GammonDD,Ota2SAQDvertCoh,AbstreiterQDMolecule06}.
The present work provides a novel route in which the potential advantages
of gate defined and self-assembled devices can be combined to yield
a scalable system. An implementation of the proposed scheme would
enable a wide range of new applications. Optical access provides a
fast spin (and charge) read-out scheme, as is desirable for implementations
of quantum error correction schemes~\cite{Taylor_architecture}.
 Robust spin-photon entanglement provides new realizations for quantum
repeaters for long-distance communication~\cite{ZollerRepeater,LiliRepeater}
and distributed quantum computing, where gate-defined quantum dots
constitute quantum computers with a limited number of qubits, which
are then connected using optical channels.

\begin{figure}
\begin{center}\includegraphics[%
  width=85mm,
  keepaspectratio]{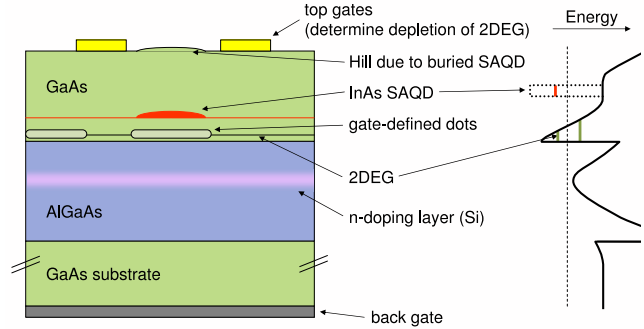}\end{center}

\caption{\label{fig:Design}Heterostructure coupling gate-defined and optically
active dots. An {}``inverted'' 2DEG is formed in the GaAs layer
using standard techniques. However, the growth of the GaAs layer will
be interrupted to grow InAs SAQDs. The surface of the sample deforms
above a SAQD, which allows alignment of the top gates used to define
lateral quantum dots in the 2DEG. The separation of the 2DEG and the
SAQDs is on the order of $10\:\mathrm{nm}$, allowing capacitative
coupling and/or electron tunneling. The conduction band energies when
moving along the growth direction are shown on the right; the energy
of the SAQD is shown as dotted line.}
\end{figure}

\emph{Design}. We envision the following structure (see Fig.~\ref{fig:Design})
containing both a 2DEG (type II) and a SAQD (type I). It consists
of an {}``inverted'' GaAs/AlGaAs heterostructure~\cite{AbstreiterInv}
with an $n$-doping layer roughly $100\,\mathrm{nm}$ below the GaAs
top layer, where due to the conduction band offset the electrons move
to the GaAs layer but remain in the vicinity of the interface  and
thereby create a 2DEG. It is possible to grow such inverted structures
with high mobilities \cite{Inv2DEGPfeiffer}. Also, metallic top gates
are grown, which can be charged negatively to deplete the 2DEG and
to create the gate-defined (lateral) quantum dots in the 2DEG. Furthermore,
during growth of the GaAs layer, another standard method is used:
addition of InAs leads to the formation of SAQDs. These dots lead
to a deformed surface at the top of the heterostructure, which allows
locating the SAQD~\cite{ImamQDCavity07} and alignment of the top
gates. 

\emph{Capacitative Coupling}. Let us now consider application of this
structure for spin read-out of the gate-defined dots. To this end,
let us assume capacitative coupling between the gate-defined dots
and the SAQD, but no tunneling takes place. We can first convert the
spin information of the electron state in the gate-defined dots to
a charge state, say, converting a triplet into a $\left(1,1\right)$
double-dot charge configuration and the singlet into a $\left(0,2\right)$
configuration \cite{PettaDD,ELreadout}. Due to the capacitative coupling
and DC Stark-shift, the excitonic transition frequency in the SAQD
will shift depending upon the charge configuration, allowing state-dependent
(i) differential transmission or (ii) photoluminescence under resonant
excitation and detection. 

We now estimate how the electric field from a single electron in the
gate defined quantum dot shifts the transitions of the optically active
quantum dot. As function of a back-gate voltage, DC stark shifts of
the exciton and charge exciton transitions of $\approx1\:\mu$eV/(kV/cm)$^{2}$
are observed~\cite{warburton02}; including a bias field of order
20 kV/cm, the effect from a small change in field is $\approx20\:\mu\mathrm{eV}/(\mathrm{kV}/\mathrm{cm})^{2}$.
In our scenario, the presence of an additional electron in our gate-defined
dot leads to such a small additional field. Assuming an oblate electron
wave function of transverse diameter $d\approx25\:\textrm{nm}>z$,
where $z$ is the distance to the self-assembled quantum dot, the
electric field along the $z$ direction is  $\approx e/2\pi\varepsilon d^{2}\approx2\:\mathrm{kV}/\mathrm{cm}$
 and we estimate the change in the DC Stark effect due to the presence
of a single electron in the gate-defined quantum dot to be a $\delta\sim10-100\mu\mathrm{eV}$
shift of the transition energy.  Consider resonance of the optical
transition when the gate-defined dot is in charge state $(1,1)$,
thus the transition rate for charge state $(0,2)$ will be suppressed
by $\alpha=(\Gamma/\delta)^{2}$, where $\Gamma$ is the homogeneously
broadened linewidth. For example, using method (ii) with detection
efficiency $\eta=1\%$, radiative decay rate $\gamma=(1\,\mathrm{ns})^{-1}$,
and measuring for $1\,\mu\mathrm{s}$, on resonance one observes $N=10$
photons on average. Taking $\alpha<1/400$ and interpreting observation
of zero or one photons as charge state $(0,2)$, the charge measurement
fails with probability $\max\{ e^{-N}(1+N),\:1-e^{-\alpha N}(1+\alpha N)\}<0.1\%$
\cite{ELreadout}. Using a cavity would speed up the read-out by the
Purcell factor. 






\emph{Adiabatic Electron Transfer}. We now consider a scheme that
relies upon electron tunneling between the SAQD and the gate-defined
dot and show that it is possible to preserve the coherence of the
pseudo-spin state. It was demonstrated experimentally that charge
can be transferred between two coupled dots made out of a quantum
well structure, using a vertical distance of 3-10$\:\mathrm{nm}$~\cite{PiTaruchaVDD01}.
Furthermore, elastic charge transfer between two vertically aligned
SAQD has been demonstrated and tunnel couplings $\tDD\approx5\:\mathrm{meV}$
and $\tDD\approx1.5\:\mathrm{meV}$ were found for dot spacings of
$5\:\mathrm{nm}$ and $10\:\mathrm{nm}$, respectively \cite{Ota2SAQDvertCoh}.
 Finally, tunneling from InGaAs SAQDs to an $n$-doped GaAs substrate
were found to be significant for a separation of $20\,\mathrm{nm}$~\cite{abstreiter2}.
Therefore, it is reasonable to assume that in the proposed structure
the tunnel coupling between the gate-defined dot and the SAQD can
be on the order of $\tDD\approx1\:\mathrm{meV}$. Since $\tDD$ is
is roughly the same as typical orbital level spacings of gate-defined
quantum dots, adiabatic transport with respect to the charge degree
of freedom can be achieved with sub-ns gate pulses. When the charge
transfer is completed, the tunnel coupling will not limit the spin
lifetime, because its effect is suppressed by the detuning $\DE$
of the discrete energy levels in each dot. Most importantly, as we
show below, \emph{an adiabatic charge transfer preserves coherence}
\emph{of the spin state}.

\begin{figure}
\begin{center}\includegraphics[%
  width=85mm,
  keepaspectratio]{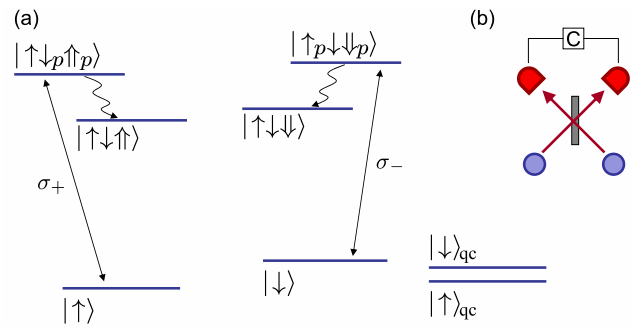}\end{center}

\caption{\label{fig:Levels}(a) The energy spectrum and the transitions of
the states on the SAQD and of the gate defined quantum computer, with
quantization axis along the growth direction. The lower states correspond
to a single electron with spin $\sigma=\uparrow,\downarrow$ on the
SAQD, $\ket{\sigma}$, and on the gate defined quantum dots, $\ket{\sigma}_{\mathrm{qc}}$.
The higher states are the charged exciton ground states, with holes
and electrons in $s$-states, $\ket{\uparrow\downarrow\Uparrow}$
and $\ket{\uparrow\downarrow\Downarrow}$. Excitation to the even
higher lying orbital $p$-states, followed by fast orbital relaxation
allows avoiding effects of the stray light of the excitation laser.
(b) Two SAQDs (circles) in remote devices each emits a photon which
is entangled with a local electron spin. These photons are then interfered
at a beamsplitter and detected---in case of a simultaneous {}``click''
at each detector the two electron spins become entangled with each
other.}
\end{figure}

We now show that the electron spin transferred between the gate defined
and the optically active dot can be used to create decoherence-protected
spin-photon entangled states. We assume that there is a magnetic field
along the growth direction such that the selection rules couple the
spin state to the circular polarization of perpendicularly emitted
photons~\cite{Cerletti05}.  First, the SAQD is prepared in the
state $(\spup+\spdown)/\sqrt{2}$, which can be achieved by preparing
this state in the gate-defined dot~\cite{DelftESR} and loading it
into the SAQD. Then, the initial SAQD state is excited into the trion
state $(\ket{\!\uparrow\downarrow\Uparrow}+\ket{\!\uparrow\downarrow\Downarrow})/\sqrt{2}$
using polarized light.  To reduce effects of the excitation laser's
stray light, one can use blue-shifted lasers to excite the electrons
and holes into the higher-lying $p$-state of the envelope wave function,
see Fig.~\ref{fig:Levels}(a). Then, these orbital states rapidly
decay in to the $s$-state; the state of the electrons relaxes to
its singlet ground state, making cycling transitions possible~\cite{lai05}.
Because the emitted phonons do not depend on spin, coherence of the
spin state is conserved. Finally, recombination occurs on a time scale
$\tRecomb$ and a photon will be emitted with a circular polarization
depending on the hole state. (Even though the radiative decay occurs
spontaneously, the total energy does not change due to the recombination
process and there is no random phase factor in the final state consisting
of an electron \emph{and} a photon.) Thus, the system is now in the
state $\ket{\EPRdotPhoton}=(\spup\ket{\sigma_{+}}+\spdown\ket{\sigma_{-}})/\sqrt{2}$,
i.e., the electron spin state on the SAQD and the photon are entangled.

To entangle the spin state of two remote SAQDs $i=1,2$, a quantum
teleportation scheme can now be applied on the state $\ket{\EPRdotPhoton}_{1}\ket{\EPRdotPhoton}_{2}$
\cite{DuanScale04,briegel98,dur99,ZollerRepeater,LiliRepeater}. 
As illustrated in Fig.~\ref{fig:Levels}(b), the photons can be interfered
at a beam splitter and, if both photons have the same polarization
and frequency, bunch together and propagate to the same detector,
so no coincidence is found. On the other hand, if the photons have
different polarization/frequency, they do not interfere and with probability
$\frac{1}{2}$ arrive at different detectors. Thus, if coincidence
is found, the electron spins on the SAQDs are projected onto the state
$\spup_{1}\spdown_{2}+\spdown_{1}\spup_{2}$, i.e., the spins are
now entangled; otherwise, the procedure is repeated. Finally, the
electrons are transferred to the nearby gate-defined quantum dots---because
spin coherence is preserved, the goal of entangling two remote (gate
defined) quantum dots is achieved.

\emph{Noise and Decoherence.} We now discuss the dominant sources
of errors. Consider the coherent spin transfer between a gate-defined
dot and a SAQD. The spin degree of freedom in each dot is defined
as the pseudo-spin of the Kramers doublet, which is split by a magnetic
field. Because SAQDs are usually grown by using a material that is
lattice-mismatched with the substrate, the two quantum dots are in
different materials and thus have different microscopic wave functions---for
example, the wave function depends on the spin-orbit coupling at the
core atoms of the crystal. The populations of the pseudo-spin states
can be transferred between the dots if it is adiabatic on the time
scale of the (minimal) Zeeman splitting. While the Zeeman splitting
$\ZeemanSplitting$ is typically much smaller than the tunnel coupling
(for InAs SAQDs $\left|g\right|\approx1$ \cite{Ota2SAQDvertCoh}
and thus $\ZeemanSplitting=50\:\mu\mathrm{eV}$ at $1\:\mathrm{T}$)
this still allows for sub-ns spin transfer. On the other hand, if
the effective spin-orbit interaction is weak (or if it is taken into
account as coherent spin rotation), different spin states do not mix
and one can still transfer the electrons on time scales of $1/\tDD$.
However, besides spin populations, spin \emph{coherence} must be preserved
as well. Because the g-factor is generally position dependent, fluctuations
in the electron position lead to spin dephasing.

To estimate this dephasing due to charge fluctuations and identify
the requirements that the spin transfer is coherent by considering
the low energy physics, it is sufficient to consider the four dimensional
Hilbert space, defined by the charge state $\ket{L}$ or $\ket{R}$
that describe if the electron is on the gate-defined dot or in the
SAQD, and by the pseudo-spin $\spup$ or $\spdown$ in that location.
We assume that the quantization axis of the Zeeman splitting is the
same in both dots, but allow for a difference  $\ZeemanDiff=\ZeemanSplitting^{L}-\ZeemanSplitting^{R}$
in the magnitude of these splittings. The system is then described
by the Hamiltonian \begin{equation}
H=\tDD\tau_{x}+\frac{1}{2}\left[\DE(t)+\noiseF(t)\right]\,\tau_{z}-\frac{1}{4}\ZeemanDiff\tau_{z}\sigma_{z},\label{eq:HnoiseModel}\end{equation}
with Pauli matrices $\sigma_{i}$ and $\tau_{i}$, acting on the spin
and charge degree of freedom, resp. Here, $\tDD$ is the tunneling
amplitude between the two dots and $\DE(t)$ is the detuning between
the states $\ket{L}$ and $\ket{R}$. We omitted the term $-\frac{1}{4}\sigma_{z}\,(\ZeemanSplitting^{L}+\ZeemanSplitting^{R})$,
because this term only leads to a trivial extra phase since $\sigma_{z}$
is conserved. Finally, we assume that the environment couples to the
charge degree of freedom by the small fluctuating force $\noiseF(t)$
and we neglect low-frequency noise. Assuming a Gaussian bath, the
charge dephasing rate due to these fluctuations is $\dephasingRate=2\int ds\:\left\langle \noiseF(s)\,\noiseF(0)\right\rangle $
\cite{thesis_Marquardt}, where $\left\langle \cdot\right\rangle $
denotes ensemble averaging. 

For the adiabatic charge transfer from $\ket{L}$ to $\ket{R}$, $\DE(t)$
is increased sufficiently slowly, such that the charge state is always
in the ground state $\ket{\psi}=\alpha\ket{L}+\beta\ket{R}$ with
respect to the spin-independent Hamiltonian. In a Born-Oppenheimer
approximation, we project the Hamiltonian on that lowest charge state
and find the coupling to the spin, $\bra{\psi}\tau_{z}\sigma_{z}\ket{\psi}=-\DE(t)\left[4\tDD^{2}+\DE(t)^{2}\right]^{-1/2}\sigma_{z}-4\tDD^{2}\left[4\tDD^{2}+\DE(t)^{2}\right]^{-3/2}\,\noiseF(t)\sigma_{z}+O\left(\noiseF^{2}\right)$.
The first term corresponds to the deterministic time evolution, which
is independent of the noise and thus does not lead to dephasing. We
focus on the second term, describing the influence of noise due to
the environment. The fluctuations $\noiseF(t)$ will lead to decoherence,
namely, a superposition $(\spup+\spdown)/\sqrt{2}$ evolves into $(\spup+e^{i\phi}\spdown)/\sqrt{2}$
with random phase $\phi=-\frac{1}{2}\int_{0}^{t}dt'\:\ZeemanDiff\left\langle \sigma_{z}\right\rangle $.
The infidelity $\errorSpinDephasing$ of the spin transfer, given
as the probability that the phase information is destroyed, is $\errorSpinDephasing\leq\frac{1}{2}\left\langle \phi^{2}\right\rangle $,
using that the bath is Gaussian. If the correlations in $\noiseF(t)$
decay on a shorter timescale than changes of $\DE(t)$ and using that
$\left\langle \noiseF(t')\,\noiseF(t'')\right\rangle $ is time-translation
invariant, we obtain \begin{align}
\errorSpinDephasing & \lesssim\frac{\dephasingRate\ZeemanDiff^{2}}{16}\,\int_{-\infty}^{\infty}dt\:\left(\frac{4\tDD^{2}}{\left[4\tDD^{2}+\DE(t)^{2}\right]^{3/2}}\right)^{2}.\label{eq:phiSqApprox1}\end{align}
Evaluating Eq.~(\ref{eq:phiSqApprox1}) for standard Landau-Zener
transition $\DE(t)=\alpha t$, we find $\errorSpinDephasing\lesssim3\pi\dephasingRate\ZeemanDiff^{2}/64\alpha\tDD$.
Taking $\tDD^{2}/\alpha\approx10$ (leading to a non-adiabatic contribution
$e^{-10}<10^{-4}$), we find the infidelity of spin transfer, \begin{align}
\errorSpinDephasing & \lesssim\frac{3}{8}\,\left(\frac{\ZeemanDiff}{\tDD}\right)^{2}\,\frac{\Gamma_{\varphi}}{\tDD}.\label{eq:infidelityVal}\end{align}
With typical parameters as used above, $\ZeemanDiff,\,\Gamma_{\varphi}\ll\tDD$,
thus the infidelity is $\errorSpinDephasing<10^{-4}$ and spins can
be transferred coherently between gate-defined dots and SAQDs.

\newcommand{\fidelity}{Fffffffffffff}

\newcommand{\errorProb}{p}

\newcommand{\successProb}{s}

\newcommand{\tauMeasure}{\tau_{\mathrm{m}}}
The optical entanglement scheme of two spin states is conditional,
i.e., it can only be applied with a certain success probability, which
is mainly determined by the efficiency $\eta$ of detecting an emitted
photon: coincidence is obtained with probability $\eta^{2}/4$. The
scheme can be repeated many times until eventually a coincidence in
the photodetectors is obtained; then the spin entanglement has error
probability $p$, which might be result from one of the following
processes. (1) From the excitonic state, a forbidden transition can
occur (e.g., $\ket{\!\uparrow\downarrow\Uparrow}\to\spdown$) with
some interconversion probability $\errorProb_{1}$ relative to the
allowed transition, destroying the desired spin-photon entanglement.
Hole spin flips would also destroy entanglement and we include these
in $\errorProb_{1}$. In Ref.~\onlinecite{lai05}, polarization preservation
of up to $90\%$ was found, even in the presence of co-tunneling and
electron-hole exchange. Because these limiting processes are not present
here, we may expect $\errorProb_{1}\leq1\%$. (2) The orbital $s$-
and $p$-states are exposed to a different hyperfine fields $\omega_{\mathrm{hf}}$,
leading to a relative phase factor of the orbitally excited states
and possibly to a spin-dependent energy of the emitted phonons. However,
because the orbitally excited state are so short-lived ($\tau_{\mathrm{ph}}\sim10\:\mathrm{ps}$),
the phonon energy is broadened and thus spin-independent; thus the
resulting error $\errorProb_{2}\sim(\omega_{\mathrm{hf}}\tau_{\mathrm{ph}})^{2}$
is negligible. (3) The excited electron could tunnel into the gate-defined
quantum dot if the excitation energy (typically a few tens $\mathrm{meV}$)
overcomes the detuning $\DE$ of the ground states; however such an
inelastic process into a discrete state is very slow \cite{inelasticFujisawa}
and so $\errorProb_{3}<10^{-3}$. (4) Spin-orbit coupling of the hole
of the excitonic state can cause decoherence, which for low temperatures
is limited by the hole spin flip time \cite{BulaevLoss_HoleDecoh}
included in $\errorProb_{1}$, thus $\errorProb_{4}<\errorProb_{1}$.
(5) For the interference at the beam splitter, photons with the same
polarization must have the same frequency $\omega$. Even if the two
remote SAQDs have different $g$-factors resulting from size fluctuations
during growth, the external static magnetic fields can be chosen such
that the Zeeman splittings are equal. However, if the hyperfine field
is not controlled, this still can lead to a frequency difference of
$\Delta\omega$. If the coincidence measurement detects photons in
a time interval $\tauMeasure$, the error is $\errorProb_{5}=(\Delta\omega\,\tauMeasure)^{2}$
\cite{LiliRepeater}. A shorter $\tauMeasure$ increases the fidelity
by broadening the photons in frequency, but decreases the detection
probability by a factor of $1-e^{-\tauMeasure/\tRecomb}$. For a typical
$\Delta\omega=(10\,\mathrm{ns})^{-1}$ and $\tauMeasure=1\:\mathrm{ns}$,
we get $\errorProb_{5}=1\%$. Finally, this allows us to estimate
the error of the entanglement of two spins to be $2\errorProb_{1}+\dots+2\errorProb_{4}+\errorProb_{5}\lesssim5\%$.

We conclude by noting new avenues opened by this work. Our approach
allows for optical interconnects between gate-controlled quantum dot
systems. The proposed structure allows using standard techniques for
integrating the SAQD in a photonic-band gap cavity. Extensions of
the present ideas may also allow reversible quantum state exchange
between spin and photon systems. These techniques enable applications
in quantum communication, such as implementing standard purification
protocols that require several qubits per optical register~\cite{briegel98,dur99},
and new architectures for scalable quantum computation using many
optically connected few-qubit registers to form a large-scale quantum
computer. Finally, for an arbitrary geometry formed in the 2DEG via
the top gates, electrons can be extracted into SAQDs placed nearby,
which allows for local measurements of single spins. For example,
when a current flows in the 2DEG, one can use the SAQD to probe if
there is spin polarization at lateral edges due to the spin Hall effect.

We thank A. Badolato, B.I. Halperin, C.M. Marcus, P. Maletinsky, A.S.
S{\o}rensen, M. Stopa, and A. Yacoby for discussions. This work was
supported by ARO/DTO, NSF Grants No. DMR-05-41988 and No. PHY-01-17795,
NSF Career award, Pappalardo Fellowship, the Harvard Center for Nanoscale
Systems, and the Packard Foundation.

\clearpage

\end{document}